\documentclass[twocolumn,prc,showpacs,superscriptaddress,preprintnumbers,amsmath,amssymb]{revtex4}
\usepackage{graphicx}
\usepackage{dcolumn}
\usepackage{bm}

\begin{document}

\preprint{}

\title{Centrality and system size dependence of (multi-strange) hyperons at 40$A$ and 158$A$ GeV: A comparison between a binary collision and a Boltzmann + hydrodynamic hybrid model.}

\author{Hannah Petersen}
\affiliation{Frankfurt Institute for Advanced Studies (FIAS),
Ruth-Moufang-Str.~1, D-60438 Frankfurt am Main,
Germany}
\affiliation{Institut f\"ur Theoretische Physik, Goethe-Universit\"at, Max-von-Laue-Str.~1, 
D-60438 Frankfurt am Main, Germany}

\author{Michael Mitrovski}
 \affiliation{Frankfurt Institute for Advanced Studies (FIAS),
Ruth-Moufang-Str.~1, D-60438 Frankfurt am Main,
Germany}
\affiliation{Institut f\"ur Theoretische Physik, Goethe-Universit\"at, Max-von-Laue-Str.~1, 
D-60438 Frankfurt am Main, Germany}

\author{Tim Schuster}
 \affiliation{Frankfurt Institute for Advanced Studies (FIAS),
Ruth-Moufang-Str.~1, D-60438 Frankfurt am Main,
Germany}

\author{Marcus Bleicher}
\affiliation{Institut f\"ur Theoretische Physik, Goethe-Universit\"at, Max-von-Laue-Str.~1,
 D-60438 Frankfurt am Main, Germany}

\date{\today}

\begin{abstract}
We present results on the centrality and system size dependence of (multi-strange)hyperons in Pb+Pb collisions at 40$A$ and 158$A$ GeV from the Ultra-relativistic Quantum Molecular Dynamics (UrQMD-v2.3) model and a coupled Boltzmann+hydrodynamics calculation. The second approach is realized in a hybrid fashion based on UrQMD with an intermediate hydrodynamical evolution for the hot and dense stage of the collision. This implementation allows a comparison of microscopic transport calculations with hydrodynamic simulations to explore the transition from a system that is not fully equilibrated such as C+C or Si+Si collisions, to a supposedly fully equilibrated system, such as that created in Pb+Pb reactions. The results of our calculation are compared to measurements of the (anti-)hyperon yields at midrapidity ($\mid$y$\mid$ $\leq$ 0.5) and total multiplicities performed by the NA49 and NA57 Collaborations at 40$A$ and 158$A$ GeV. Furthermore, we compared our predictions to the centrality dependence of $\Lambda$, $\bar{\Lambda}$ and $\Xi^{-}$ rapidity spectra and total multiplicities at 40$A$ and 158$A$ GeV, where possible.  
 \end{abstract}


\maketitle

Heavy ion collisions offer the possibility to study nuclear matter under extreme conditions, at high temperatures and densities, in the laboratory. This can be achieved by colliding nuclei, either by shooting accelerated ions at a stationary target, such as at the Alternating Gradient Synchrotron (AGS) at the Brookhaven National Laboratory (BNL) or the CERN Super Proton Synchrotron (SPS), or by head-on collisions of two ion beams, such as at the Relativistic Heavy Ion Collider (RHIC) at the BNL and the Large Hadron Collider (LHC) at CERN. To achieve a ``macroscopic" volume of excited nuclear matter, very heavy nuclei such as lead (Pb) or gold (Au) are used. At high temperature and baryon density, nuclear matter is expected to melt into a state of free quarks and gluons, known as the quark gluon plasma (QGP)~\cite{Bass:1998vz,Wang:1996yf,Harris:1996zx}. 

The strange ($s$) and anti-strange ($\bar{s}$) quarks are a well-suited diagnostic tool to investigate the properties of the matter created in heavy ion reactions, because they are newly produced and not contained in the colliding nuclei. The final state distributions of strange hadrons reflect the strangeness production in the hot and dense stage of the reaction. Strangeness is a good quantum number because it is conserved in strong interactions. Strange quarks (and therefore strange hadrons) decay via the weak interaction, where typical lifetimes are on the order of $10^{-10}$ s and so these decays are not altered on the timescale of a hadronizing QGP and can be well reconstructed by experiments. 

In Refs.~\cite{Rafelski:1982pu,Koch:1986ud} the authors suggested almost 25 years ago that strange particle production is enhanced in the QGP with respect to a hadron gas and serves as a unique signature of the quark gluon plasma. This enhancement can be quantified relative to a reaction in which a transition to a QGP phase does not take place, such as p+p collisions for which the system size is very small. The enhancement occurs, because in the QGP different channels are responsible for strangeness production, leading to a strong reduction of the strangeness equilibration time as compared to the hadronic scenario. This is supported by the difference in threshold energies due to the fact that in a deconfined state, only the strange quarks have to be produced, rather than strange hadrons themselves. 

\begin{table}
\caption{\label{tab:table1} The energy threshold to create strange hadrons in hadronic interactions.}
\begin{ruledtabular}
\begin{tabular}{lc|}
Production reaction &$E_{\rm{threshold}}\ (\rm{MeV})$\\
\hline
$N + N \rightarrow \Lambda + K^{+} + N$ & $\approx$ 700\\
$N + N \rightarrow N + N + \Lambda + \bar{\Lambda}$ & $\approx$ 2200 \\
$N + N \rightarrow N + N + \Xi^{-} + \bar{\Xi}^{+}$ & $\approx$ 4500 \\
$N + N \rightarrow N + N + \Omega^{-} + \bar{\Omega}^{+}$ & $\approx$ 5200 \\
\end{tabular}
\end{ruledtabular}
\end{table}

\begin{figure*}
\includegraphics[scale=0.9]{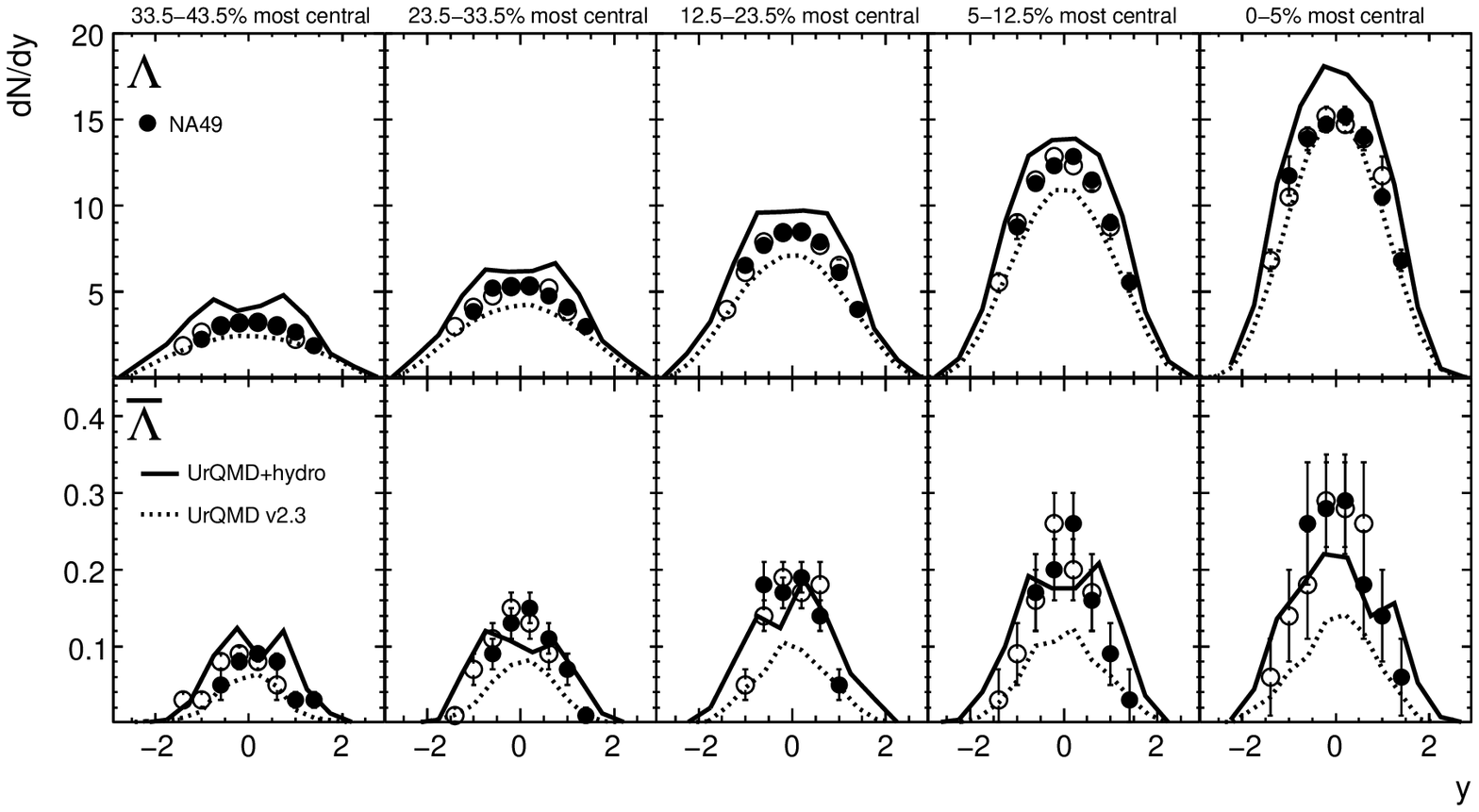}
\caption{\label{fig:fig1}Rapidity distribution of $\Lambda$ (top) and $\bar{\Lambda}$ (bottom) from NA49~\cite{Anticic:2009ie} measured for different centralities in Pb + Pb collisions at 40$A$ GeV. The closed symbols indicate measured points, whereas the open points are reflected with respect to midrapidity. The solid line represents calculations with UrQMD + hydro and the dashed line represents calculations with UrQMD.}
\end{figure*}

\begin{figure*}
\includegraphics[scale=0.8]{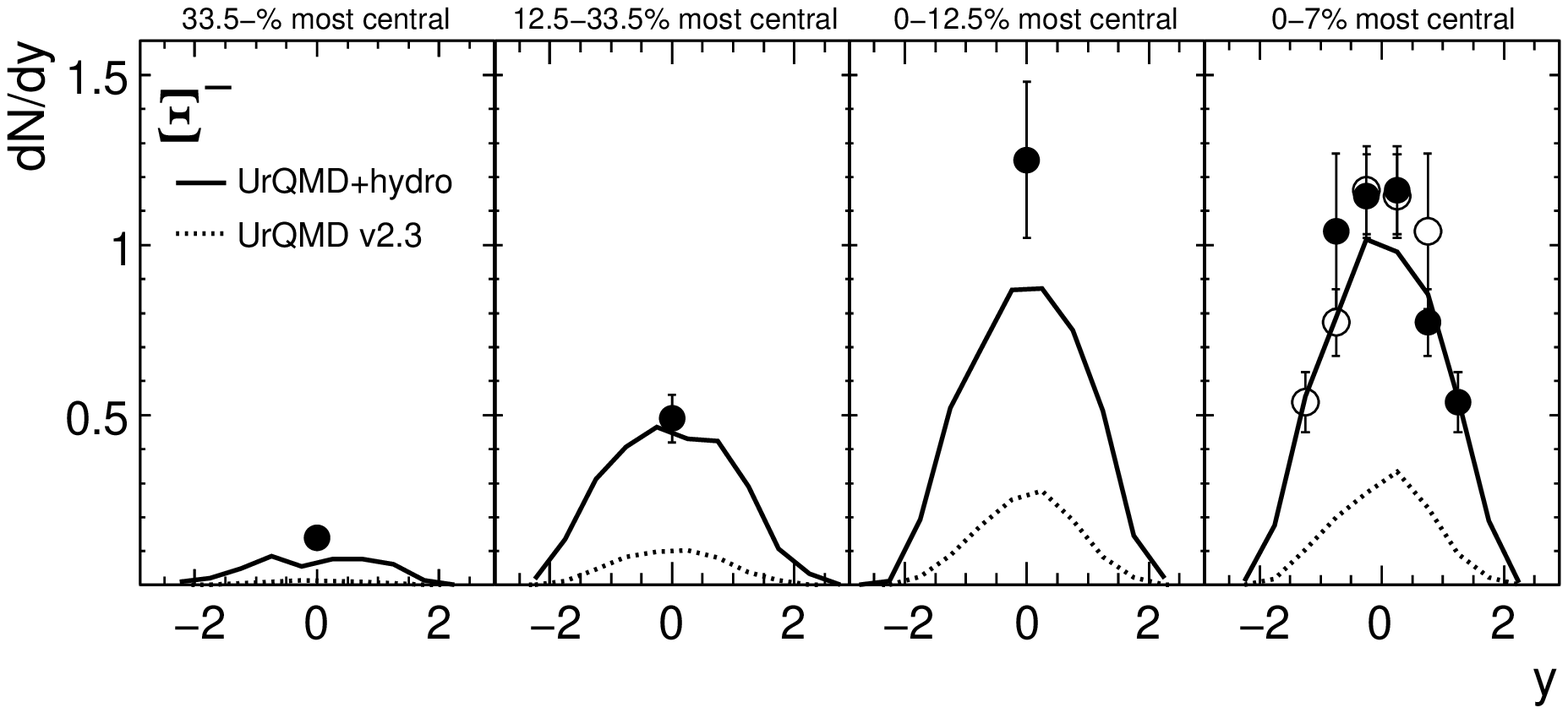}
\caption{\label{fig:fig2}Rapidity distribution of $\Xi^{-}$ from NA49~\cite{Anticic:2009ie,Alt:2008qm} measured for different centralities in Pb + Pb collisions at 40$A$ GeV. For the unmeasured rapidity spectra we present  the measured yield at midrapidity ($\mid$y$\mid$ $\leq$ 0.5). The solid line represents calculations with UrQMD + hydro and the dashed line represents calculations with UrQMD.}
\end{figure*}

Hadronic transport approaches without a QGP phase transition usually have problems in describing the abundances of (multi-)strange baryons. In a hadronic interaction, a high threshold energy is needed to create strange hadrons, which is calculated from the differences in masses between the initial and final state particles, as shown in Table~\ref{tab:table1} for N + N $\rightarrow$ (anti-)hyperons. Because of these large threshold energies (multi-)strange baryon production is suppressed even in initial collisions in the SPS energy regime ($\sqrt{s_{\rm{NN}}} \approx$ 5-20 GeV)~\cite{Drescher:2001hp,Bleicher:2001nz}. A further important source for (multi-strange)hyperon production are multistep processes that lead to high mass resonance excitations and the rescatterings of initially produced antikaons. In a hadron resonance gas it takes some time to equilibrate strangeness and the timescale of a heavy ion collision might be shorter than this equilibration time. In regions of high net baryon density, the production of strange baryons is more likely, especially $\Lambda$s produced in association with a kaon because this reaction has a rather low threshold energy. Therefore the $\Lambda$ yields are usually reproduced also in nonequilibrium hadronic cascade approaches. 

Different approaches to enhance strangeness production in semihadronic scenarios without a transition to a deconfined state have been proposed. Two different types of models explain how an enhanced production of (anti-) hyperons can be understood. The first model \cite{Vance:1999pr} applies a baryon junction exchange mechanism, whereas the other model uses multihadronic interactions that bring the system to local chemical equilibrium~\cite{Greiner:2000tu}. A similar way to push the system into chemical equilibrium is with multiple interactions of pions~\cite{Cassing:2001ds} or the oversaturation of pion number~\cite{Rapp:2000gy}. Calculations in Ref.~\cite{BraunMunzinger:2003zz} show that reaching an equilibrium in a pure hadronic medium, considering collision rates and timescales of the hadronic fireball expansion, is only possible if hadronic multiparticle interactions are taken into account. Recently, counterarguments against this conclusion were suggested within a hadronic model that includes Hagedorn states~\cite{NoronhaHostler:2007jf}. 

\begin{figure*}
\includegraphics[scale=0.9]{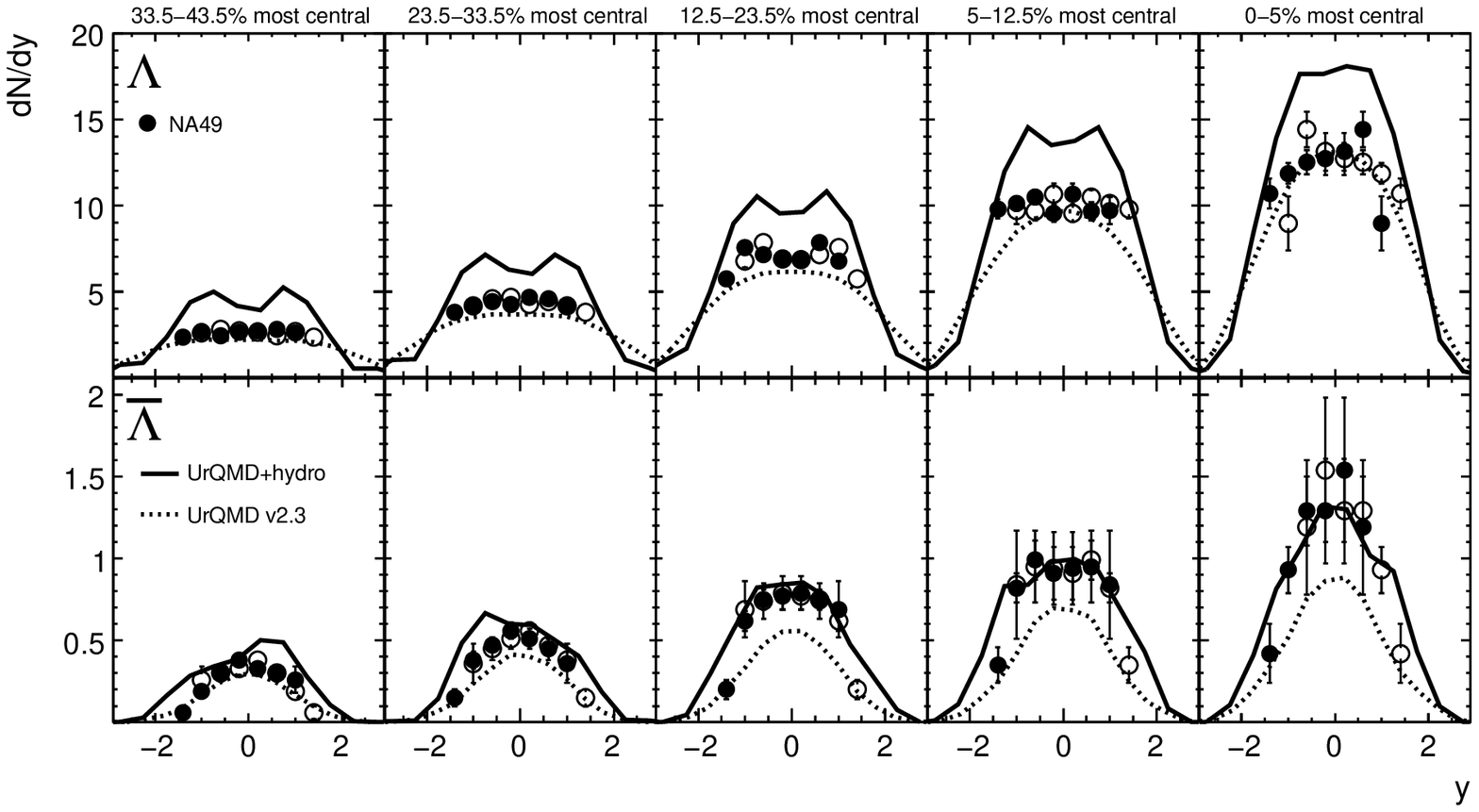}
\caption{\label{fig:fig3}Rapidity distribution of $\Lambda$ (top) and $\bar{\Lambda}$ (bottom) from NA49~\cite{Anticic:2009ie} measured for different centralities in Pb + Pb collisions at 158$A$ GeV. The closed symbols indicate measured points, whereas the open points are reflected with respect to midrapidity. The solid line represents calculations with UrQMD + hydro and the dashed line represents calculations with UrQMD.}
\end{figure*}

\begin{figure*}
\includegraphics[scale=0.83]{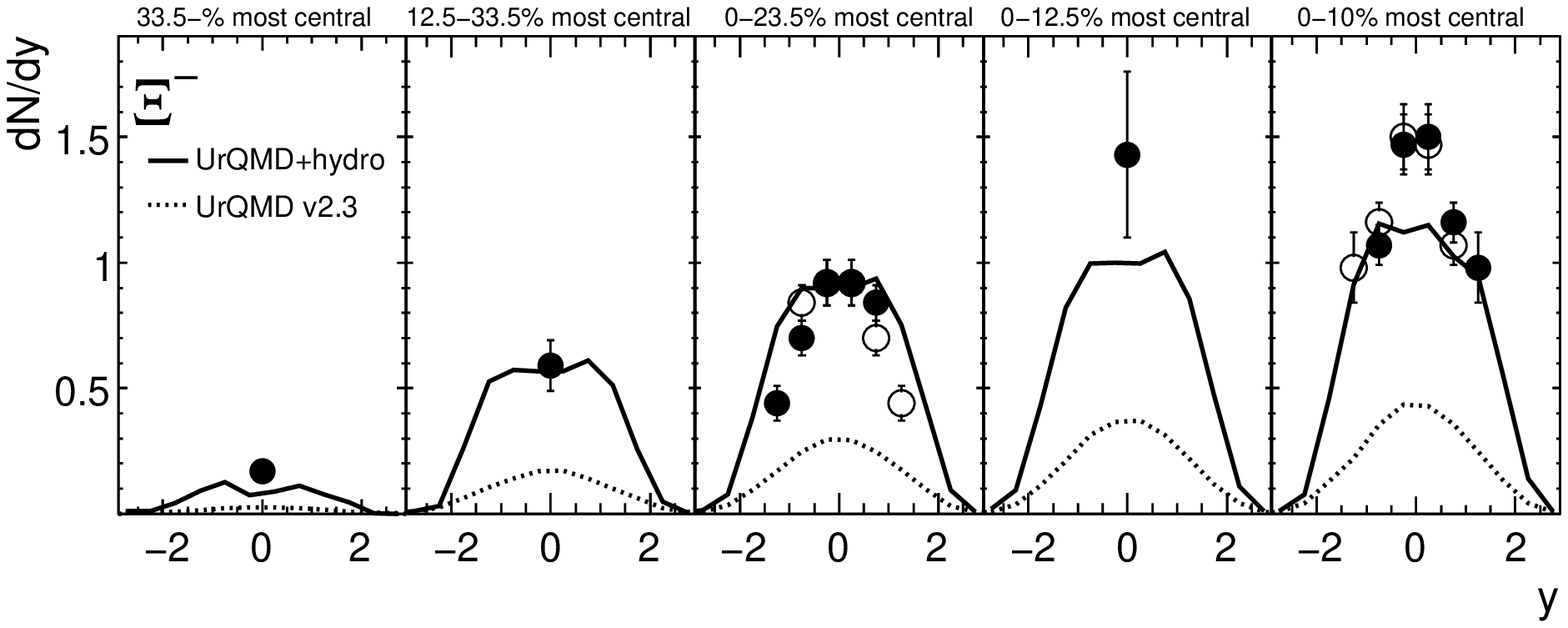}
\caption{\label{fig:fig4}Rapidity distribution of $\Xi^{-}$ from NA49~\cite{Anticic:2009ie,Alt:2008qm,Mitrovski:2007zz} measured for different centralities in Pb + Pb collisions at 158$A$ GeV. For the unmeasured rapidity spectra we present  the measured yield at midrapidity ($\mid$y$\mid$ $\leq$ 0.5). The solid line represents calculations with UrQMD + hydro and the dashed line represents calculations with UrQMD.}
\end{figure*}

The apparent enhancement of strangeness (or more precisely a lifting of the suppression) can be explained also in statistical models by the transition from a canonical ensemble, for small systems (e.g. like in p + p), which treat charge conservation  exactly, to a grand canonical ensemble for heavier systems like Pb + Pb of Au + Au. A variation of the statistical model assuming a fully thermalized core where strangeness is produced according to phase-space distribution surrounded by a corona which can be described as a superposition of elementary p-p reactions has been proposed to explain the system size and centrality dependence of strangeness production in heavy ion collisions \cite{Aichelin:2008mi,Becattini:2008ya}. The equilibrium assumption seems to be a necessary requirement to explain the strange particle yields. However, none of the preceding approaches allowed for a consistent description of both spectra and multiplicities of (multi-)strange baryons up to now.  

\begin{figure}
\includegraphics[scale=0.5]{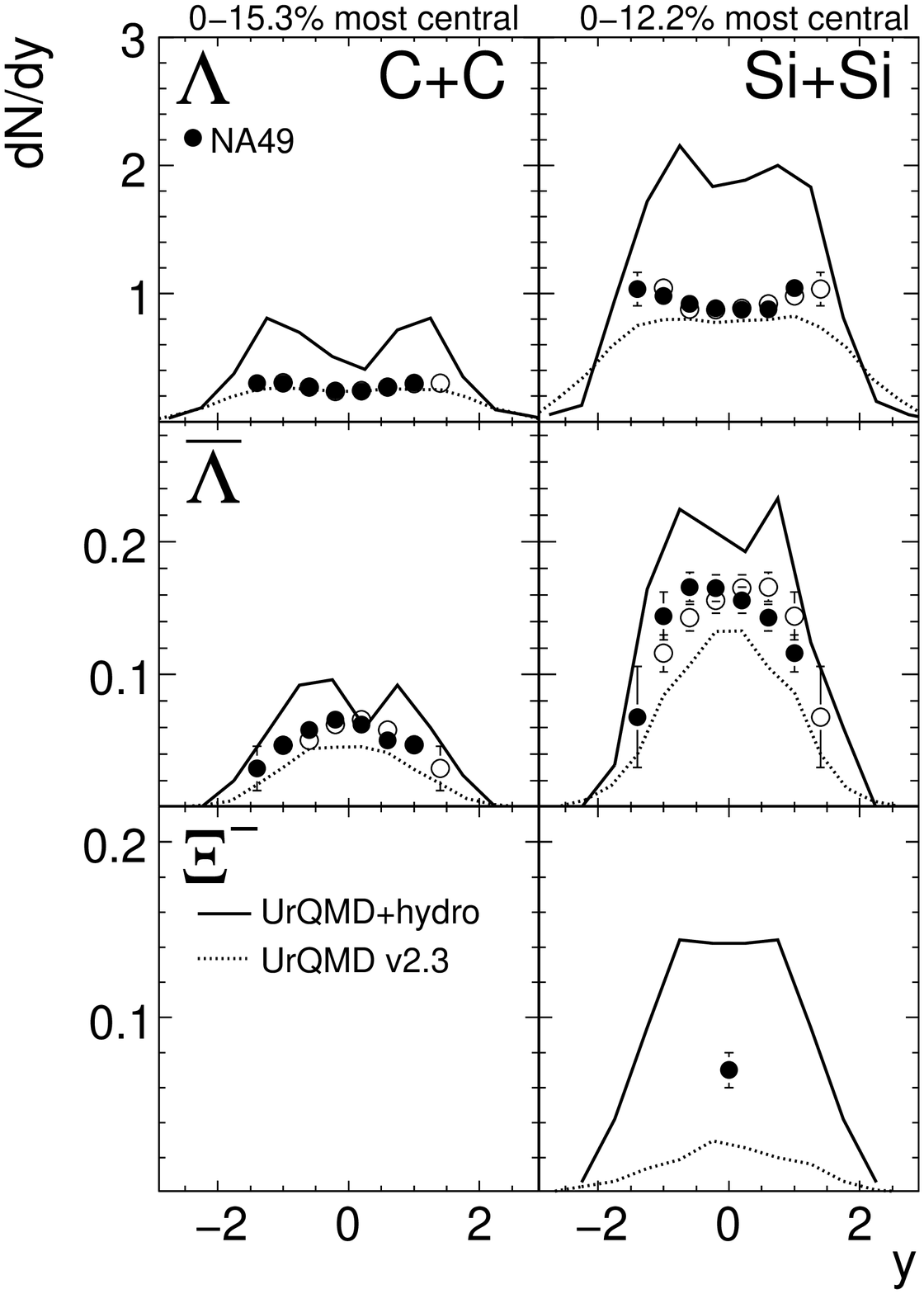}
\caption{\label{fig:fig5} The rapidity spectra of $\Lambda$, $\bar{\Lambda}$ and $\Xi^{-}$ produced in central C + C (left) and Si + Si (right) collisions at 158$A$ GeV~\cite{Anticic:2009ie,Alt:2004wc}. The closed symbols indicate measured points: open points are reflected with respect to midrapidity. The solid line represents calculations with UrQMD + hydro and the dashed line represents calculations with UrQMD.}
\end{figure}

In this article, we attack this question by using a hadronic transport approach, namely the Ultra-relativistic Quantum Molecular Dynamics (UrQMD) in version 2.3~\cite{Bleicher:1999xi,Bass:1998ca,Petersen:2008kb}, and compare the results to a hybrid approach in which a hydrodynamic evolution is embedded into the intermediate stage of the model dynamics \cite{Petersen:2008dd}. By doing this, we do not aim to understand the dynamics of the strangeness equilibration process itself, but we are able to compare strangeness production in a full non-equilibrium approach with a dynamical approach involving local thermal and chemical (especially also for strangeness) equilibrium. The advantage of this hybrid approach compared to a pure hydrodynamic approach is that it allows for the direct comparison between non-equilibrium and locally equilibrated dynamics and accounts for a realistic decoupling of the particles with separate chemical and kinetic freeze-outs because the late stage rescattering is included.  

UrQMD is a microscopic many body approach to p + p, p + A and A + A interactions at relativistic energies and is based on the covariant propagation of color strings, constituent quarks and diquarks accompanied by mesonic and baryonic degrees of freedom. The model includes rescattering of particles, the excitation and fragmentation of color strings and the formation and decay of hadronic resonances. Detailed information about the latest version of the UrQMD model can be found in Ref.~\cite{Petersen:2008kb}. 

In the hybrid model, UrQMD serves to calculate the initial state of a heavy ion collision for the hydrodynamical evolution to account for the non-equilibrium effects in the very early stage of the collision~\cite{Steinheimer:2007iy}. With this procedure event-by-event fluctuations are naturally included. Furthermore, the baryon density fluctuations in one event are incorporated which might lead to higher strangeness production as discussed in Ref.~\cite{Steinheimer:2008hr}. The initial conditions for calculations at different centralities and for different system sizes are also provided by the UrQMD approach without further parameter adjustment. The spectators are propagated during the whole evolution in the hadronic cascade, whereas the other particles are mapped to the hydrodynamical grid. Then, the full (3 + 1) dimensional hydrodynamic evolution proceeds employing the SHASTA algorithm (SHarp And Smooth Transport Algorithm) ~\cite{Rischke:1995ir,Rischke:1995mt}. For the present comparison we restrict ourselves to a hadron gas equation of state which includes the same degrees of freedom as in UrQMD. The strangeness chemical potential is adjusted in the calculation of the equation of state in such a way that the net-strangeness vanishes locally. In the transport approach the local net strangeness might take a finite value, but this happens mostly due to the latestage rescatterings. The initial strangeness fluctuations and the influence on the strangeness production are further discussed in Ref.~\cite{Steinheimer:2008hr} and subject to future investigations.

\begin{figure}
\includegraphics[scale=0.5]{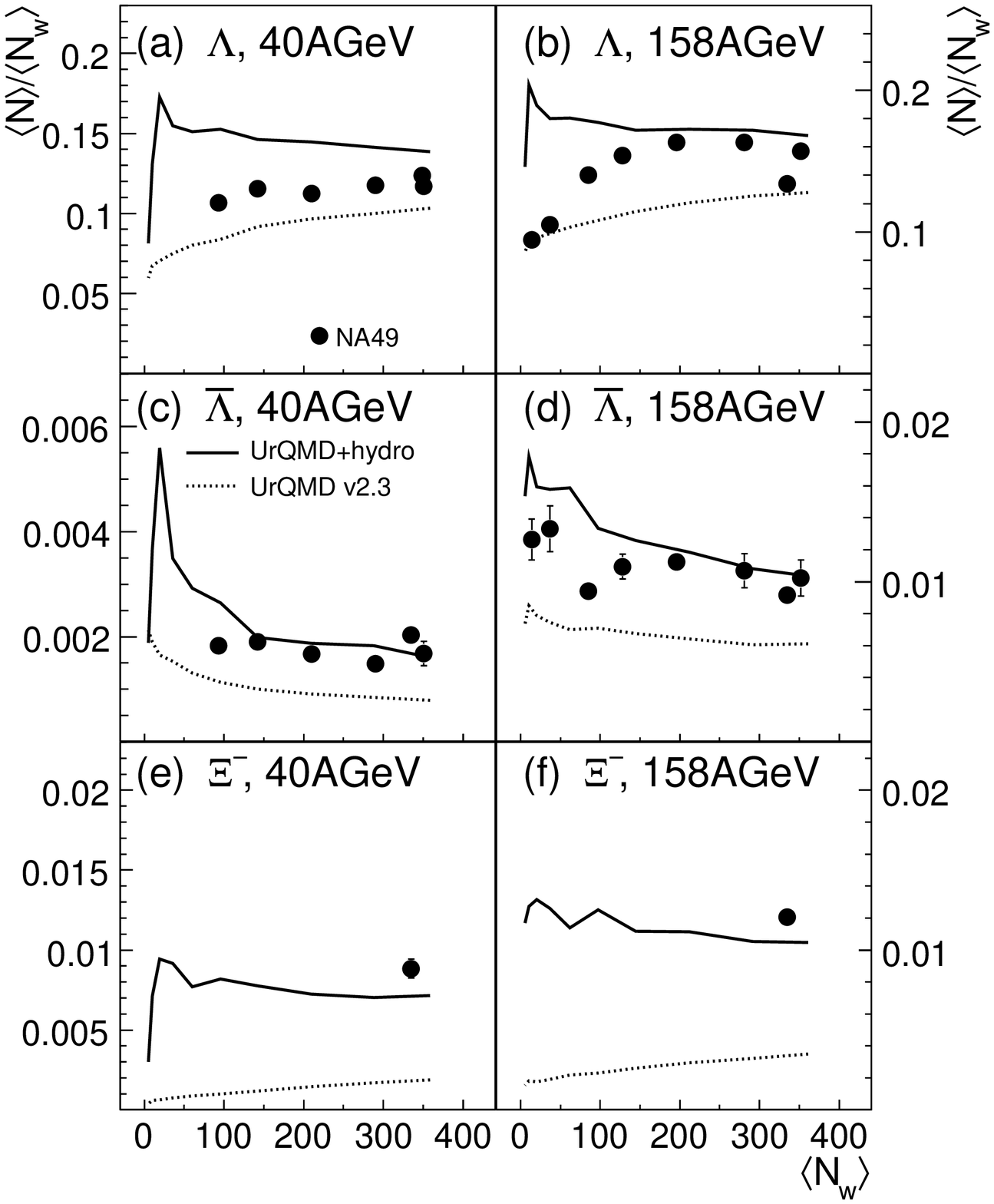}
\caption{\label{fig:fig6} The points correspond to the measured total yield by NA49 at 40$A$ (left) and 158$A$ GeV (right)~\cite{Anticic:2009ie,Alt:2008qm,Alt:2004wc,Alt:2004kq}. The solid line represents calculations with UrQMD + hydro and the dashed line represents calculations with UrQMD.}
\end{figure}

The transition from the hydrodynamic phase to the Boltzmann dynamics, when the system is sufficiently diluted, takes place when the energy density $\epsilon$ of a transverse slice of cells drops below 5 times the ground state energy density $\epsilon_{0}$. This gradual transitive procedure provides a reasonable approximation to an iso-eigentime criterion and has been applied successfully in Refs.~\cite{Li:2008qm,Petersen:2009vx}. This switching criterion corresponds to a T $-$ $\mu_{B}$-configuration in which the chemical freeze-out is expected, approximately, at T = 170 MeV for $\mu_{B}$ = 0. The hydrodynamic fields are then mapped to particle degrees of freedom via the Cooper-Frye equation on an isochronous hypersurface in each individual slice. For the production of strange baryons the corresponding chemical potentials $\mu_{B}$ and $\mu_{S}$ are taken into account. The subsequent rescatterings and final decays are calculated in the UrQMD model. 

Let us stress again that the present hybrid approach uses only a free hadron gas equation of state and has no phase transition included. More details about this approach including multiplicities and spectra can be found in Ref.~\cite{Petersen:2008dd}. For a general comparison to similar approaches the reader is referred to Ref.~\cite{Dumitru:1999sf,Teaney:2001av,Nonaka:2006yn,Hirano:2007ei,Hama:2009pk}. The great advantage of this approach is that the whole dynamical evolution of the heavy ion reaction is generated, and it allows us, therefore, to investigate many observables, such as flow \cite{Petersen:2009vx}, HBT radii \cite{Li:2008qm}, and transverse mass spectra \cite{Petersen:2009mz,Steinheimer:2008hr} at the same time and under the same conditions. 

In the following we show results from UrQMD and the hybrid approach (UrQMD + hydro) both employing only hadronic degrees of freedom. We compare these results to the measured total particle yields~\footnote{Note that only the NA49 experiment has the acceptance to measure the total multiplicity.} and the abundances at midrapidity from the NA49 and NA57 experiments in Pb+Pb collisions. Furthermore, we compare the model predictions to the measured $\Lambda$, $\bar{\Lambda}$ and $\Xi^{-}$ rapidity spectra in minimum bias Pb + Pb and central C + C and Si + Si collisions from the NA49 Collaboration at 40$A$ and 158$A$ GeV. In addition, we present calculations from UrQMD and UrQMD + hydro for the centrality dependence of $\Lambda$, $\bar{\Lambda}$ and $\Xi^{-}$ rapidity spectra at $E_{\rm lab}=40A$ GeV and $E_{\rm lab}=158A$ GeV and compare these to the measured spectra or yields at midrapidity ($\mid$y$\mid$ $\leq$ 0.5).

Figures~\ref{fig:fig1} and~\ref{fig:fig2} show the rapidity spectra for $\Lambda$, $\bar{\Lambda}$ and $\Xi^{-}$ at $E_{\rm lab}=40A$ GeV predicted from UrQMD (dashed lines) and UrQMD + hydro (solid lines) at different centrality bins and the data are presented as full circles (note that the open points represent the reflected points at midrapidity)~\cite{Anticic:2009ie,Alt:2008qm}. We concentrate here on multiplicity distributions represented by rapidity spectra to explore the centrality dependence of particle yields. UrQMD predicts a lower multiplicity compared to the hybrid approach. This can be understood, because at the transition from the hydrodynamic evolution to the hadronic cascade the particles are produced according to thermal distributions. Particle production means in this context the transition from macroscopic degrees of freedom, such as energy density and net baryon number density distributions to real particle degrees of freedom as they are propagated in the microscopic approach. In this manner, it is much easier to produce especially multistrange hyperons than via resonance and string excitation. The hadronic cross sections of (multi-)strange baryons are small, so they do not get reequilibrated during the hadronic expansion phase once they are produced: for example the total number of $\Xi^-$ at $E_{\rm lab}=40A$ GeV decreases from 2.57 directly after the hydrodynamic phase to 2.35 as the final value, which is far from the pure transport value of 0.64. The only exception are the $\bar{\Lambda}$s which may get absorbed even during the late stage of the evolution. These effects during the hadronic rescattering stage are usually neglected in hydrodynamic approaches without a hadronic cascade after the Cooper-Frye freeze-out. 

The $\Lambda$ production in the pure transport approach is in reasonable agreement with the experimental data because the production threshold is not as high as for the other hyperons because they are likely to be produced in nucleon-nucleon collisions in association with a kaon (see Table~\ref{tab:table1}). Therefore the $\Lambda$s reflect the initial baryon stopping and the rapidity distribution of the net baryon number. In the hybrid model calculation the $\Lambda$ production is higher because strangeness is chemically equilibrated. We overpredict the Lambda production in the hybrid approach by 10$-$20 \% which is similar to the deviations of statistical model fits to Lambda yields, as can be seen, for example in Ref.\cite{Andronic:2008gu}. In contrast to that, the nucleons reflect the net baryon density distribution which is explicitly propagated in the hydrodynamic evolution. The assumption of chemical equilibrium leads to enhanced $\bar{\Lambda}$ and $\Xi$ production and is in a good agreement with the experimental data, which are shown in Figs. \ref{fig:fig3}-\ref{fig:fig4}~\cite{Anticic:2009ie,Alt:2008qm,Mitrovski:2007zz} for the highest SPS energy. There is no qualitative difference regarding this observation between the results from the two beam energies. This might lead to the conclusion that the dynamics for strangeness production are similar in the whole SPS energy range. The difference between UrQMD (dashed line) and UrQMD+hydro (solid line) is at $E_{\rm lab}=40A$ GeV small for $\bar{\Lambda}$'s. 

A complementary strategy to varying the centrality of the collision to explore the applicability limit of the hybrid model in terms of collision volume is to perform calculations for smaller systems. Figure~\ref{fig:fig5} represents the rapidity spectra of  $\Lambda$ and $\bar{\Lambda}$ for central C + C and Si + Si collisions at 158$A$ GeV measured by the NA49 Collaboration~\cite{Anticic:2009ie,Alt:2004wc}. Even if the number of participants (see Table~\ref{tab:table2} in the Appendix) in central Si + Si collisions is similar to peripheral Pb + Pb collisions (33.5$-$43.5 \% most central), the geometry is completely different and the starting time for the hydrodynamic evolution is different (dependence on the radius of the nucleus $R$). It is evident that for such small systems UrQMD (dashed lines) describes the data better than UrQMD + hydro (solid lines). The limited system size does not allow for the same degree of equilibration in C + C and Si+Si interactions as in the case of Pb + Pb collisions for all shown centralities. This indicates that up to Si + Si reactions, equilibration cannot be achieved and the system is merely a (non-)trivial convolution of elementary baryon-baryon (and in smaller parts meson-baryon) interactions. From this, we also conclude that the possibility of a phase transition is reduced in such small systems because the creation of a QGP would lead to higher strangeness production.  

\begin{figure}
\includegraphics[scale=0.47]{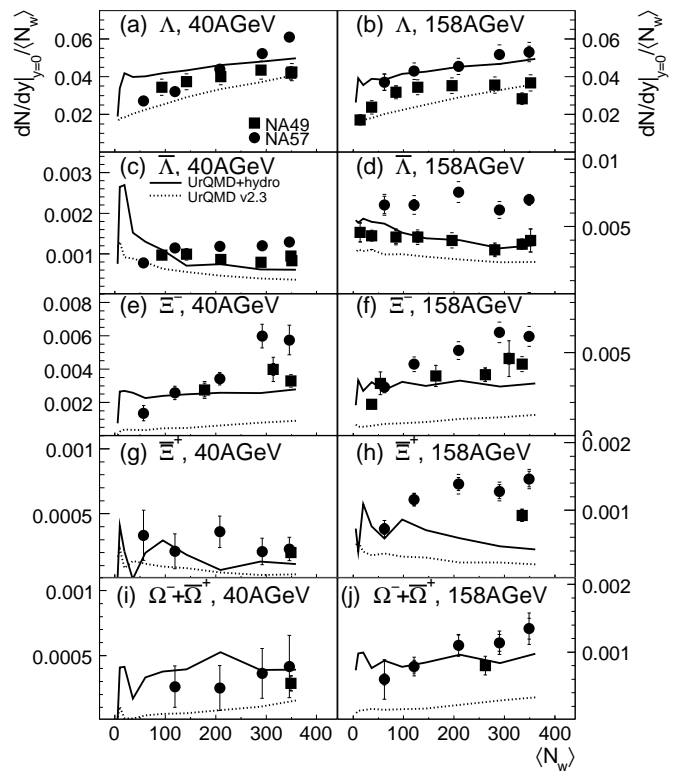}
\caption{\label{fig:fig7} Centrality dependence of the (anti-) hyperon yields at midrapidity ($\mid$y$\mid$ $\leq$ 0.5) at 40$A$ GeV (left) and 158$A$ GeV (right) measured by NA49 (squares)~\cite{Anticic:2009ie,Alt:2008qm,Alt:2004wc,Alt:2004kq} and NA57 (circles)~\cite{Antinori:2006ij,Antinori:2004ee,NA57_Webpage}. The solid line represents calculations with UrQMD+hydro and the dashed line represents calculations with UrQMD.}
\end{figure}

Let us next focus on a detailed study of the centrality dependence of strange baryons in Pb + Pb reactions. Figure~\ref{fig:fig6} depicts the predictions of the centrality dependence as a function of the number of wounded nucleons $\langle N_{W}\rangle$ of the total yield of $\Lambda$, $\bar{\Lambda}$, and $\Xi^{-}$ at $E_{\rm lab}=40A$ GeV and $E_{\rm lab}=158A$ GeV~\cite{Anticic:2009ie,Alt:2008qm,Alt:2004wc,Alt:2004kq}. One clearly observes that for central collisions the assumption of an equilibrated system leads to a suitable description of the strangeness production within the hybrid model calculation. To investigate if this interpretation holds also for smaller phase-space volumes let us turn to the particle production at midrapidity. 

The centrality and system size dependence of (anti-) hyperon production at 40$A$ and 158$A$ GeV at midrapidity ($\mid$y$\mid$ $\leq$ 0.5)  is shown in Fig.~\ref{fig:fig7} measured by the NA49~\cite{Anticic:2009ie,Alt:2008qm,Alt:2004wc,Alt:2004kq} and NA57~\cite{Antinori:2006ij,Antinori:2004ee,NA57_Webpage} Collaborations. A discrepancy between both experiments is observed. However, for a qualitative interpretation, the present data are sufficient and we will therefore refrain from speculations on the origin of the experimental inconsistencies. The dashed line represents calculations with UrQMD and the solid line depicts predictions from UrQMD + hydro simulations. The pure transport approach describes the data in very peripheral collisions quite well. This indicates that very peripheral Pb+Pb or small systems like C + C/Si + Si are not in equilibrium, as discussed earlier. Moving to more central Pb + Pb collisions, it is evident that the thermal production of strange quarks is in much better agreement with the measured data than obtained in the binary collision prescription. Especially for multistrange hyperons it is visible that the production via independent strings and resonance excitations is not sufficient to explain the experimentally measured yields. From this, we conclude that the assumption of equilibration alone allows us to explain a major part of the strangeness enhancement in central collisions of massive nuclei which does not necessarily invoke the formation of a QGP. For peripheral reactions or small collision systems a microscopic description of binary interactions between hadrons is favored.

\begin{figure}
\includegraphics[scale=0.45]{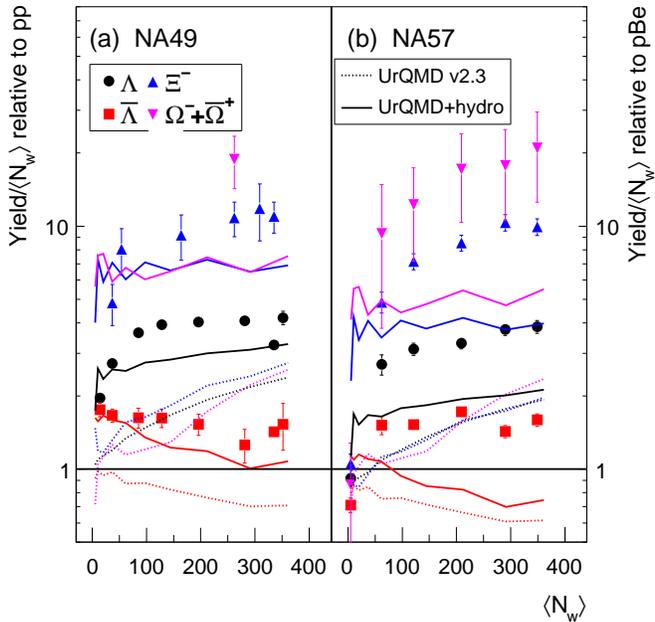}
\caption{\label{fig:fig8}(Color online) Hyperon enhancement as a function of $N_{w}$ at 158$A$ GeV measured from NA49 (a) and NA57 (b). Note that NA57 uses p + Be instead of p + p as baseline.}
\end{figure}

Finally, the hyperon enhancement as a function of the number of wounded nucleons, $\left<N_{w}\right>$, at 158$A$ GeV measured by NA49 [Fig.~\ref{fig:fig8}(a)]~\cite{Anticic:2009ie,Alt:2008qm,Alt:2004wc,Alt:2004kq} and NA57 [Fig.~\ref{fig:fig8}(b)]~\cite{Antinori:2006ij,Antinori:2004ee,NA57_Webpage}, is shown. The strangeness enhancement is defined as the yield per participant in Pb + Pb collisions normalized to the yield per participant in p + p or p + Be interactions: 

\begin{equation}
\label{eq:equation1}
E = \left( \frac{Yield}{N_{w}} \right)_{A+A}  / \left( \frac{Yield}{N_{w}} \right)_{p+p/p+Be},
\end{equation}

The centrality dependence is essentially given by the volume of the matter in the hydrodynamic stage at the transition temperature. This results in a moderate centrality dependence of the strange particle yields per participant from mid-peripheral to central reactions because the transition temperature is basically independent from the centrality (apart from modifications due to the slight change in baryon density). However, one should note that for very peripheral reactions an additional suppression might be present due for the need of a canonical statistical description~\cite{Redlich:2001kb} not employed here. The assumption of a grand canonical ensemble is always an infinite thermal bath and particle reservoir independent of centrality. For $\bar{\Lambda}$, the absorption during the hadronic rescattering stage is reflected in the centrality dependence. A clear hierarchy is visible depending on the strangeness content. The shape of the centrality dependence is changing from a decreasing distribution for $\bar{\Lambda}$ to an increasing shape for $\Lambda$, $\Xi^{-}$ and $\Omega^{-}+\bar{\Omega}^{+}$ with an enhancement for $\Omega^{-}+\bar{\Omega}^{+}$ of approximately 10. The absorption probability of $\bar{\Lambda}$ increases with centrality while for the other (multi)strange baryons the enhancement is stronger for larger collision volumes. UrQMD clearly underpredicts the effects, in all cases. 

In this article, we presented predictions for the system and centrality dependence of strange (anti-)hyperons at 40$A$ and 158$A$ GeV from the Ultra-relativistic Quantum Molecular Dynamics model (UrQMD) and a hybrid model that incorporates a (3 + 1)-dimensional hydrodynamic evolution into the UrQMD transport approach. In so doing this, a full non-equilibrium hadronic transport approach is compared to a hybrid model in which strangeness is produced according to equilibrium distributions locally. It is important to note that no explicit phase transition to the QGP has been taken into account in the present calculation. This approach takes into account event-by-event fluctuations and is able to generate different centralities and system sizes without readjusting parameters. 

We started by showing the comparison of the centrality dependence of the rapidity spectra for $\Lambda$, $\bar{\Lambda}$ and $\Xi^{-}$ in Pb + Pb collisions at $E_{\rm lab} = 40A$ GeV and $E_{\rm lab} = 158A$ GeV, where it was visible that due to the thermal production of strangeness in the UrQMD + hydro model a higher hyperon yield is generated. Compared to Pb + Pb collisions the system seemed not fully equilibrated in C + C and Si + Si collisions and therefore UrQMD provided a better description of the measured rapidity spectra. The conclusion is that strangeness equilibration is only reached in Pb + Pb collision at least in the shown centrality classes. This confirms the findings of previous studies based on statistical analysis of strange particle yields within an approach that includes proper expansion dynamics. The production of strange (anti-)hyperons in smaller systems like C + C and Si + Si can be described with a binary scattering transport approach like UrQMD. 

This work was supported by the the Hessian LOEWE initiative through HIC for FAIR, GSI, BMBF, DFG and DESY. We are grateful to the Center for Scientific Computing (CSC) at Frankfurt for the computing resources. H.~Petersen gratefully acknowledges financial support by the Deutsche Telekom Stiftung and support from the Helmholtz Research School on Quark Matter Studies. T.~Schuster gratefully acknowledges support from the Helmholtz Research School on Quark Matter Studies. The authors would also like to thank C. Blume and R. Stock.

\appendix

\section{Appendix}

\begin{table}[tbh]
\caption{\label{tab:table2}Summary of centrality which is quantified by the number of the total inelastic cross section and the average number of wounded nucleons $\langle N_{W} \rangle$ for Pb+Pb collisions at 40$A$ GeV and Pb+Pb, Si+Si and C+C collisions at 158$A$ GeV.}
\begin{ruledtabular}
\begin{tabular}{ccc}
  $E_{Beam}$           & 
  Centrality (\%)        &
  $\langle N_{W} \rangle$    \\  \hline
  
Pb+Pb 40$A$ GeV       &   0 --  5.0 & 359  \\
           &   5.0 -- 12.5 & 289  \\
           & 12.5 -- 23.5 & 209  \\
           & 23.5 -- 33.5 & 142  \\
           & 33.5 -- 43.5 &  95  \\
           & 43.5 -- 54.0 &  60  \\
           & 54.0 -- 64.0 &  36  \\
           & 64.0 -- 74.0 &  19  \\
           & 74.0 -- 84.0 &  10  \\
           &  84.0 -- 94.0 &  5  \\     \hline
            
Pb+Pb 158$A$ GeV     &   0.0 --  5.0    & 361  \\
            &   5.0 -- 12.5  &  291  \\
            &  12.5 -- 23.5 & 212  \\
            &  23.5 -- 33.5 & 144  \\
            &  33.5 -- 43.5 &  97  \\
            &  43.5 -- 54.0 &  62  \\
            &  54.0 -- 64.0 &  37  \\
	   &  64.0 -- 74.0 &  20  \\
            &  74.0 -- 84.0 &  10  \\
            &  84.0 -- 94.0 &  5  \\ \hline

Si+Si 158$A$ GeV    & 0 - 12.2 & 37 \\  \hline

C+C 158$A$ GeV    & 0 - 15.3 & 13 \\

\end{tabular}
\end{ruledtabular}
\label{tab:mbdatasets}
\end{table}

\end{document}